\documentclass[aps,prb,twocolumn,floatfix]{revtex4}
\usepackage{amsmath}

\begin{document}

\title{Separating the articles of authors with the same name}

\author{ Jos\'e M. Soler\footnote{E-mail: jose.soler@uam.es} }
\affiliation{ Departamento de F\'{\i}sica de la Materia Condensada, 
              C-III, Universidad Aut\'{o}noma de Madrid,
              E-28049 Madrid, Spain }

\date{\today}

\begin{abstract}
   I describe a method to separate the articles of different authors
with the same name.
   It is based on a distance between any two publications, defined in
terms of the probability that they would have as many coincidences
if they were drawn at random from all published documents.
   Articles with a given author name are then clustered according to
their distance, so that all articles in a cluster belong very likely
to the same author.
   The method has proven very useful in generating groups of papers
that are then selected manually.
   This simplifies considerably citation analysis when the author 
publication lists are not available.
\end{abstract}


\maketitle

   Citation analysis has become an essential tool for research 
evaluation.\cite{Moed2005}
   Generally, the evaluation referees are provided with a list of 
publications of the individuals or groups to be evaluated, although
frequently these are in a format (say, on paper) that is not easy
to use for searches in citation databases.\cite{ISI}
   Furthermore, the widespread accessibility of these databases to 
the full research community has estimulated less formal evaluations, 
in which publication lists are not available.
   In such cases, the publication lists themselves must be generated
from the databases, complementing the author names with their
affiliations and research fields.
   When even these are not well known (say, because only the last
affiliation and research field are known) the search must be based 
on the author name only.
   This poses the problem of extracting the articles of the desired
author, among those of other authors with the same name.

   In this work I address this problem by defining a distance 
between any two given articles, based on the coincidences between
them.
   This allows to cluster related articles, so that all the articles 
of a cluster are likely to belong to the same author.
   This reduces the problem to that of selecting the apropriate 
clusters, rather than each individual article.

   Distances between documents have been proposed on the basis of
coincidences of words and phrases as well as $n$-grams (sequences of
$n$ consecutive characters),\cite{Damashek1995} and these distances
have been used for a wide range of tasks, like language 
classification, or collecting documents on a given subject.
   In the present case, we are interested in relating documents whose
full text is usually not available, while their abstract is generally 
available but relatively expensive to handle in terms of database 
access and storage.
   Instead, documents are characterized by a record with a variety of 
fields, like author names and addresses, title, research field, 
keywords, journal and year of publication, etc.\cite{ISI}
   Since coincidences in all these fields are significant for 
identifying their authors, the problem arises of how to combine them
in a consistent way.
   Thus, one needs to answer questions like: are two papers `closer'
if they were published in the same journal or if they have $n$ common 
words in their titles? Or if they have a common coauthor?

   To solve this problem, I will propose the following general idea:
imagine that you draw two documents at random from the entire
database of $N_D$ documents.
   The probability that they coincide in everything (that is, that
the same document is drawn twice) is obviously $1/N_D$.
   The probability that they coincide in any given feature is also 
well defined in principle.
   For example, if $n_j$ of the documents in the database were 
published in a given journal $j$, the probability that the two random 
articles were published in that journal is $(n_j/N_D)^2$.
   The probability that the two random articles had a 
journal-of-publication coincidence {\it less or equal likely} than
that is $\sum_{i=j}^{N_J} (n_i/N_D)^2$, with the $N_J$ journals ordered 
by decreasing order of their number of articles in the database.

   Then I will define the distance $D_{ij}$ between two documents $i$ 
and $j$ by
\begin{equation}
   D_{ij} = \log_{10}( P_{ij} ) - \log_{10}( 1/N_D )
\label{Dij}
\end{equation}
where $P_{ij}$ is the probability that two random documents would have
overall coincidences less or equal likely than those between
$i$ and $j$.
   Clearly, $i=j \Rightarrow P_{ij}=1/N_D$ and $D_{ij}=0$.
   On the other extreme, if $i$ and $j$ do not coincide in anything, 
then $P_{ij}=1$ and $D_{ij} = \log_{10}(N_D)$ will be maximum.

   Obviously, $P_{ij}$ is highly nontrivial to calculate, especially
for multiple, correlated coincidences.
   However, it turns out that very crude approximations still lead to 
meaningful distances that are useful for our purposes.
   Therefore, as a first approach, I will make two extremely crude 
approximations: 
1) assume that all possible values of a given field (say author
names, like R. Smith and J. M. S. Torroja) are equally probable; and
2) ignore any correlations between different coincidences
(like address words Harvard and Massachusetts).
   I will divide each field in `words', and allow only one instance 
of each word within the field (that is, if the word Spain appears
twice in the list of author addresses, I will take it only once).
   Some words, like articles and prepositions of the
title, will be excluded.
   Thus, each field will be characterized by an estimated number 
of possible word values occurring in it.
   For example, if the estimated number of journals is $N_J$,
the approximated probability that they are equal for two random 
articles is $1/N_J$.
   More generally, if the estimated number of possible word values in 
a field is $N$, and there are $n_i$ and $n_j$ different words in that
field of articles $i$ and $j$, the probability that exactly $n_{ij}$ 
of them coincide (in any order) is
\begin{eqnarray}
\lefteqn{ p(n_{ij}|n_i,n_j,N) = } \nonumber \\
 & \frac{n_i! ~n_j! ~(N-n_i)! ~(N-n_j)!}
   {N! ~n_{ij}! ~(n_i-n_{ij})! ~(n_j-n_{ij})! ~(N-n_i-n_j+n_{ij})!}
\label{Pnij}
\end{eqnarray}
which is the probability of getting $n_{ij}$ common balls from
two independent random extractions of $n_i$ and $n_j$ balls out 
of a set of $N$ different balls.
   The probability of getting {\it at least} $n_{ij}$ coincidences
is simply 
$P(n_{ij}|n_i,n_j,N) = 1-\sum_{n=1}^{n_{ij}-1} p(n|n_i,n_j,N)$.
   Then, ignoring also correlations between different fields, 
I will approximate the distance between $i$ and $j$ by
\begin{equation}
D_{ij} \simeq \log_{10}(N_D) + \sum_{f=1}^{N_F} \log_{10}\left(
   P(n^f_{ij}|n^f_i,n^f_j,N^f) \right)
\label{Dijaprox}
\end{equation}
where $f$ indexes the $N_F$ different record fields.

   Table~\ref{fields} shows the estimated number of possible values
for the fields provided by the standard records of the ISI-Thomson
Web of Knowledge~\cite{ISI} (excluding `abstract' and `cited references').
\begin{table}
\begin{tabular}{lc}
Field & $\log_{10}$(Size) \\
\hline
Documents ($N_D$)& 8.0 \\
Author names     & 4.0 \\
Email            & 6.0 \\
Address words    & 2.0 \\
Title words      & 2.0 \\
Keywords         & 3.0 \\
Research field   & 2.0 \\
Journal          & 2.0 \\
Publication year & 1.0 \\
\hline
\end{tabular}
\caption{
   Assumed number of possible values taken by the different fields
that characterize a document record from the ISI-Thomson Web of Knowledge
database~\cite{ISI}.
}
\label{fields}
\end{table}
   Notice that most of the assumed values are much lower than the true
number of possible options.
   Rather, they are set so that $1/N$ is roughly the probability of
the most frequent word in that field (i. e. $\sim 10^{-3}$ is the 
estimated probability of an author name like R. Smith).
   Even thus, when two articles are `close' (i. e. when they belong to
the same author), the neglect of correlations implies a large 
underestimation of the probability of the combined coincidences,
making $D_{ij}$ negative.
   The important point, however, is that, when the two articles 
{\em do not} belong to the same author, the coincidences are rarely 
sufficient to make $D_{ij} < 2$, which is what one would expect for 
the probability $P_{ij} \simeq 10^2/N_D$ that two random articles belong
to the same author (assuming that the average author has published
$\sim 10^2$ articles).

   It is not unfrequent that an author changes the affiliation and,
simultaneously, the field of research (for example after finishing 
the PhD).
   Still, it is common that she/he publishes a pending work in the former 
field (and perhaps with some of the former coauthors) but using already 
the new affiliation.
   In this case, it is possible to trace the common author identity in
the two groups of apparently unrelated papers.
   To allow this, I define a new set of distances as
\begin{equation}
   d_{ij} = \min_k( D'_{ik}+D'_{kj} ), ~~~\mbox{where}~~~ 
   D'_{ij} = \max( D_{ij}, 0 )
\label{dij}
\end{equation}
where $k$ runs over all the papers with the given author name.
   A similar redefinition of distances has been proposed for nonlinear
dimensionality reduction,\cite{Tenenbaum2000,Roweis-Saul2000} where
$k$ was restricted to a small neighborhood of $i$ and $j$.
   In the present case, however, distances are strongly non Euclidian
and multidimensional scaling \cite{Mardia1979} has not proven
particularly useful.

   The problem of classifying or clustering a set of elements according
to their distances is highly nontrivial.\cite{Mardia1979}
   In our case, however, this task is facilitated by the neglection of 
correlations and the subsequent underestimation of distances between
articles of the same author, since this creates a large gap between 
these distances and those among different authors.
   In practice, I simply make clusters of papers that have 
zero distance (notice that the definition of $d_{ij}$ implies that all 
the distances among the cluster members must be zero).
   The resulting clusters of papers, generated with the values of
Table~\ref{fields}, tend to give some `false negatives' (i. e. 
different clusters that belong to the same author) but rarely
`false positives' (papers of different authors within the same cluster), 
except perhaps for the most common author names
(for these, it may be necessary to increase $N_D$, or to decrease the
other values of Table~\ref{fields}, in order to increase the distances).

   The clusters are then presented interactively (by showing one or more 
representative papers of the cluster), in different possible orders,
for their selection or rejection.
   Other clues, like the period of publication of the cluster papers,
or the distance to previously selected clusters, are also provided
to help in the selection.
   Thus, in most cases it is very obvious which clusters must be
selected, and the selection process is very fast and straightforward.
   Once the largest clusters have been considered, it is convenient
to swicth to an order of presentation by increasing distance to the
selected papers and, as soon as this distance becomes larger than
$\sim 3$,
the remaining clusters may be rejected altogether.
   This is important since, most generally, the main inconvenience is 
the large number of small clusters (many of them with a single paper)
that apparently belong to different authors.
   The following shows the begining of the selection dialog for
a typical case of intermediate complexity (for the name of this author,
Soler JM):
\begin{widetext}
\begin{center}
\begin{verbatim}
Found   142 papers in   18 groups
group, papers, citations =   1    99    4364
group, papers, citations =   2    20     207
group, papers, citations =   3     1     130
group, papers, citations =   4     2      36
group, papers, citations =   5     3      34
group, papers, citations =   6     3      18
group, papers, citations =   7     2       5
group, papers, citations =   8     1       3
group, papers, citations =   9     1       3
group, papers, citations =  10     1       2
group, papers, citations =  11     1       1
group, papers, citations =  12     1       1
group, papers, citations =  13     1       0
group, papers, citations =  14     1       0
group, papers, citations =  15     1       0
group, papers, citations =  16     1       0
group, papers, citations =  17     2       0
group, papers, citations =  18     1       0

Group    1 has    99 papers and   4364 citations in period 1981-2006
Distance to selected groups is ******   A sample paper is
 Title: Density-functional method for very large systems with LCAO basis sets
 Authors:  SanchezPortal, D; Ordejon, P; Artacho, E; Soler, JM;
 Source:  Int. J. Quantum Chem. (1997) 65, 453:461
 Address words: AUTONOMA MADRID FIS MAT CONDENSADA E-28049 SPAIN 
   NICOLAS CABRERA OVIEDO E-33007
 Select this group? (y|n|u|all|none|p|c|d|(number)|help):
\end{verbatim}
\end{center}
\end{widetext}
   The first group of papers is mine, without any false positives.
   In this case there are neither false negatives (i. e. none
of the papers in the other groups are mine), although this is not
the most usual case.

   In summary, a practical algorithm has been presented for separating
the papers of an author from those of other authors with the same name.
   It semi-automates the separation process by creating clusters of
papers that most likely belong to the same author, thus simplifying
greatly the generation of an author publication list.
   
   I would like to acknowledge very useful discussions with
J. V. Alvarez, R. Garc\'{\i}a, J. G{\'o}mez-Herrero, L. Seijo, and 
F. Yndurain.
   This work has been founded by Spain's Ministery of Science through
grant BFM2003-03372.

\appendix

\section*{Appendix: How to get and process an ISI-Thomson SCI file}

   In order to find in practice the merit indicators of an author, 
one can follow these steps:
\begin{enumerate}
\item
   Download the programs {\it filter} and {\it merit} from this
   author's web page,\cite{soler_web} and compile them if necessary.
\item
   Perform a ``General search'' in the ISI-Thomson Web of Science 
   database~\cite{ISI} for the author's name. Appropriate filters
   may be set already in this step, if desired.
\item
   Select the records obtained. Usually the easiest way is to check
   ``Records from 1 to last\_one'' and click on ``ADD TO MARKED LIST''
   (if you find too many articles, you may have to mark and 
   save them by parts, say (1-500)$\rightarrow$file1, 
   (501-last\_one)$\rightarrow$file2);
\item
   Click on ``MARKED LIST''.
\item
   Check the boxes ``Author(s)'', ``Title'', ``Source'', ``keywords'',
   ``addresses'',  ``cited reference count'', ``times cited'',
   ``source abbrev.'', ``page count'', and ``subject category''.
   Do not check ``Abstract'' nor ``cited references'', since this
   would slow down considerably the next step.
\item
   Click on ``SAVE TO FILE'' and save it in your computer.
\item
   Click on ``BACK'', then on ``DELETE THIS LIST'' and ``RETURN'',
   and go to step 2 to make another search, if desired.
\item
   Use the {\it filter} program to help in selecting the papers
   of the desired author.
   Mind for hidden file extensions, possibly added by your navigator, 
   when giving file names in this and next step.
\item
   Run the {\it merit} program to find the merit indicators.
\end{enumerate}

\bibliographystyle{apsrev}
\bibliography{separation}

\end{document}